# Correction of systematic errors in scanning tunnelling spectra on semiconductor surfaces: the energy gap of Si(111)-7x7 at 0.3 K

S. Modesti [1,2,3], H. Gutzmann [3], J. Wiebe[3], and R. Wiesendanger[3]

[1] TASC-INFM National Laboratory, ss 14 km 163.5, I-34149 Trieste, Italy
[2] Physics Department, University of Trieste, v. Valerio2, I-34127 Trieste, Italy
[3] Institute of Applied Physics and Microstructure Research Center, University of Hamburg, Jungiusstrasse 11, D-20355 Hamburg, Germany.

The investigation of the electronic properties of semiconductor surfaces using scanning tunnelling spectroscopy (STS) is often hindered by non-equilibrium transport of the injected charge carriers. We propose a correction method for the resulting systematic errors in STS data, which is demonstrated for the well known Si(111)-(7x7) surface. The surface has an odd number of electrons per surface unit cell and is metallic above 20 K. We observe an energy gap in the ground state of this surface by STS at 0.3 K. After correction, the measured width of the gap is (70 ± 15) meV which is compatible with previous less precise estimates. No sharp peak of the density of states at the Fermi level is observed, in contrast to proposed models for the Si(111)-(7x7) surface.

Some of the most exciting recent discoveries in condensed matter research involve low dimensional electron systems [1]. The electronic structure of the ground state of these systems and of nanostructures can be successfully studied with atomic resolution by scanning tunnelling spectroscopy (STS) at low temperatures [2]. When the low-dimensional system or the nanostructure is integrated on a semiconductor surface the effects of the non-equilibrium transport of charge carriers from the surface just below the tip to the contact at the metallic sample holder can induce appreciable distortions in the spectra at low temperature. These effects may be large even at room temperature on wide-band-gap semiconductors [3].

STS is based on the relationship between the tunnelling current I and the bias voltage $V_{tip-surf}$ applied between the metallic tip of a scanning tunnelling microscope (STM) and the sample surface below it. This is only a fraction of the measured bias $V_{bias}$ that is applied between the STM tip and the metallic sample holder, if the sample is not a metal. The potential drop $V_{surf-holder}$ between the sample surface and the sample holder is caused by the penetration of the tip electric field into a region below the sample surface (tip-induced band bending [4]) and by the additional potential drops caused by the current in the bulk and in the depletion regions that often occur at the surface and at the contact between the sample and the sample holder. The latter three potential drops are strongly temperature and current dependent, get large at low temperature when the density of the minority carriers vanishes, and induce systematic errors in STS that are often not corrected.

A clear example of these current-induced effects is given by the results on the Si(111)-(7x7) surface by Myslivecek et al. [5] showing a ~1 eV gap in the STS spectrum at 7 K. This apparent energy gap increases with decreasing resistance $R_{total}=V_{bias}/I$ between tip and sample holder, which can be adjusted by changing the tip to surface distance. The energetic positions of the other features of the spectrum also shift by several tenths of an electronvolt as a function of $R_{total}$, pointing to a charge transport effect [5,6].

The Si(111)-(7x7) surface is metallic at room temperature, having an odd number of electrons per unit cell. At low temperature, a gap in the ground state can open when the electron-correlation is strong enough to overbalance the electron kinetic energy. This drives the system into a Mott-Hubbard state. Another possible reasons for a gap at $E_F$ are structural distortions such as charge density waves. There is no theoretical estimation of the Mott-Hubbard or charge density wave gap for the Si(111)-(7x7) surface, and the experimental data from high-resolution photoemission [7], high-resolution electron-energy-loss spectroscopy [8], nuclear magnetic resonance [9] and transport measurements [10,11] hint to a gap between 0 eV and a few tenths of an eV at low temperatures. These values are well below the minimum gap reported from the published STS data. In this paper we propose and test a method for the correction of the systematic errors caused by the transport of the carriers injected by the STM tip and evaluate the gap of the ground state of the Si(111)-(7x7) surface at 0.3 K.

Let us consider a semiconductor surface and assume that the surface gap is small or absent. This is valid for many semiconductor surfaces, but not for the clean (110) surfaces of III-V semiconductors. If the surface gap is small the pinning of the position of the Fermi level $E_F$ with respect to the surface states is effective. Then, in the limit of vanishing tunnelling current, the surface band bending induced by the small gap is negligibly dependent on the presence of the STM tip, on its distance from the surface, and on the applied bias $V_{bias}$ between the tip and the sample holder [12].

A nonzero current induces an additional potential drop when it crosses the depletion region below the surface –if present-, the bulk of the sample, and the contact region between the sample and the sample-holder. The fractions of the potential drops in the bulk and at the holder-sample contact are independent of the tip-surface distance and depend only on the value of the current for fixed sample and holder properties. For effective pinning of the Fermi level $E_F$ at the surface the fraction of the potential drop in the region immediately below the surface is also independent of the tip-surface distance and of the applied $V_{bias}$. Therefore, the total potential drop between the surface and the sample holder $V_{surf-holder}$ depends only on the value of the tunnelling current (see Fig. 1).

Thus, a good approximation of the dependence of $V_{surf-holder}$ on the tunnelling current I can be obtained by measuring the tunnelling spectra with the largest and smallest possible overall resistance values $R_{total}$. If $R_{total}$ is small and thus comparable to the resistance between the surface and the sample holder, $V_{tip-surf}$ is negligible with respect to $V_{surf-holder}$ and $V^s_{bias}(I)$ – the total bias measured at small $R_{total}$ for a given I - is a good approximation of $V_{surf-holder}(I)$ (see Fig. 1 and 2). For a spectrum recorded at large $R_{total}$ the relationship between $V_{tip-surf}$ and I can be obtained by $V_{tip-surf}(I)= V^l_{bias}(I) - V^s_{bias}(I)$, where $V^l_{bias}(I)$ is the total bias measured at large $R_{total}$. In order to verify whether this simple method works we have measured STS spectra of the Si(111)-(7x7) surface as a function of temperature, resistance $R_{total}$, and of the contact resistance between the sample and the holder.

We used heavily n (As) doped Si(111) wafers with a resistivity of less than 0.005 Ω cm which were mounted on Ta or Mo holders. In one set of measurements in an STM system operating between 0.3 K and 40 K [13] the samples were cleaned by flashing up to 1600 K by electron bombardment onto the back of the sample. In a second set taken in an STM system operating between 5 K and 300 K the flashes were obtained by direct current heating. We used clean W tips, and the reproducibility of the results was tested with Pt-Ir tips. The bias voltage $V_{bias}$ is applied to the sample holder. $I(V_{bias})$ and $dI/dV_{bias}(V_{bias})$ spectra were recorded with open feedback after feedback-stabilizing the tip-sample separation at a resistance of $R_{total}= V_{stab}/ I_{stab}$ using a stabilization current and voltage of $I_{stab}$ and $V_{stab}$, respectively. The tip-sample distance was taken constant during the acquisition of the spectra. The differential conductance is either measured directly by Lock-In technique using a small bias modulation $V_{mod} < 10mV$ (1.6 kHz) or by numerical differentiation of the $I(V_{bias})$ curves.

A set of $dI/dV_{bias}$ spectra measured at 0.3 K on a well ordered (7x7) surface for different $R_{total}$ is shown in Fig. 3. Each spectrum is the average of 200 spectra measured at different points of the surface unit cell visible in the inset constant current topograph. The gap and all the other features in the spectra shift as a function of $R_{total}$ as described in Ref. [5]. The spectrum for $R_{total}=130$ GΩ has a gap of 0.35 eV and rest atom and adatom features are shifted in energy by several tenths of eV [5]. Similar shifts are measured at 4 K, but with increasing the temperature to 30 K (spectra not shown) the shifts are disappearing. This observation is in agreement with trends discussed in Ref. [5]. According to the hypothesis stated above the true $V_{tip-surf}(I)$ at $R_{total}=130$ GΩ can be approximated by the difference between $V_{bias}(I)$ measured at $R_{total}=130$ GΩ and that measured at $R_{total}=800$ MΩ (see Fig. 1). By using the resulting relations between $V_{tip-surf}$ and I and $V_{tip-surf}$ and $V_{bias}$ we multiply the lock-in measured $dI/dV_{bias}(V_{bias})$ by $dV_{bias}/dV_{tip-surf}$, plot it as a function of $V_{tip-surf}$ and divide by $I(V_{tip-}$

$_{\text{surf}}$)/$V_{\text{tip-surf}}$ to obtain the correct (dI/dV$_{\text{tip-surf}}$)/(I/V$_{\text{tip-surf}}$). The (dI/dV$_{\text{tip-surf}}$)/(I/V$_{\text{tip-surf}}$) and dI/dV$_{\text{tip-surf}}$ curves are shown in Fig. 4a and 4b together with the linear fits of the near gap regions used to evaluate the gap. Our estimation of the energy gap is (70 ± 10) meV. The energies of the other features in the spectrum, corresponding to the rest atoms (R; -0.8 eV) and adatoms (A; -0.15 eV and 0.3 eV) contributions to the surface density of states, are in agreement with the energies measured at 295 K (Fig. 4c and Refs. [5] and [14]), a temperature where the charge transport effects should be negligible. The peak positions are also in agreement with those measured by high resolution photoemission spectroscopy [7]. The normalized conductivity spectrum obtained by correcting the spectrum recorded with R$_{\text{total}}$=27 GΩ in Fig. 4a shows a smaller gap and a small shift of the features of the empty states. This occurs because V$_{\text{bias}}$(I) recorded for R$_{\text{total}}$=800 MΩ contains a small contribution from V$_{\text{tip-surf}}$, and the errors caused by this contribution increase when the gap resistance of the two curves used to obtain the corrected spectrum get more similar. Due to the same reason, the estimated value of the energy gap of (70 ± 10) meV is a lower limit of the real energy gap.

The comparison of the R$_{\text{total}}$=130 GΩ spectrum with the room temperature and the photoemission data suggests that the correction described above is valid. A second test can be done by minimizing the contribution of the current dependent shifts caused by the contact between the sample and the sample holder. We have done this by flashing the samples at high temperature by direct current heating in a second STM system that operates between 5 K and 300 K. The high current density and temperature reached at the contact area between the Si surface and the Ta clips that clamp the wafer improve the electrical conductance of the contacts at low temperature. The resulting normalized conductivity (dI/dV$_{\text{bias}}$)/(I/V$_{\text{bias}}$) measured in this second run at 5 K and with R$_{\text{total}}$=80 GΩ (Fig. 4c,d) shows an energy gap comparable to that of the corrected spectrum (dI/dV$_{\text{tip-surf}}$)/(I/V$_{\text{tip-surf}}$) of Fig. 4a and similar energies for the rest atom (R) and adatom (A) features are found. These energies stay constant even if R$_{\text{total}}$ is decreased by a factor 10 (not shown) or if the temperature is risen from 5 K to 295 K (fig.3c). The absence of appreciable temperature and R$_{\text{total}}$ dependent effects at high gap resistances in this set of spectra indicate that most of the shifts in Fig. 3 are mainly caused by the contact between the sample and the sample-holder. The changes in the relative intensities of the rest atom and adatom features are caused by differences in the electronic structure of the STM tips.

The value for the energy gap estimated at 5 K from the second data set of Fig 4c,d is (70 ± 15) meV. This value is an upper limit of the real gap since a possible residual contribution from the potential drop between the surface and the sample holder may result in a V$_{\text{tip-surf}}$ smaller than V$_{\text{bias}}$. Since the spectra have been measured at constant tip-surface distance, the sensitivity of dI/dV$_{\text{bias}}$ spectra is limited to about 1/20$^{\text{th}}$ of the dI/dV$_{\text{bias}}$ value measured at the minimum at E$_F$ at 295 K. Therefore our estimation of the gap is based on a linear fit of the density of states above and below the gap down to a density 0.05 times lower than the minimum observed at room temperature, and about 2% of the density of states 0.3 eV above or below E$_F$. We are not sensitive to states with a density lower than this limit. The lower and upper limits of the gap from the corrected 0.3 K data and from the 5 K data are compatible within the experimental error. Although the two values are not measured at the same temperature, it is likely that the change in the energy gap between 0.3 and 5 K is within our experimental error because the gap opens between 15 K and 20 K (from temperature dependent spectra, data not shown). Our best estimate of the energy gap of the ground state of Si(111)-(7x7) is therefore (70 ± 15) meV.

Let us now compare this value with the other available experimental data. The high resolution photoemission data by Barke et al. [7] indicate a metallic surface at 20 K. But, Fig. 2a of this reference shows that the density of states (DOS) at $E_F$ decreases appreciably from 50 K to 20 K, falling to at least half of its initial value. We observe the opening of the gap in the STS spectra between 15 K and 20 K, where the minimum of the differential conductivity at $E_F$ reaches the zero level within our experimental error. However, we caution the reader about the fact that the value of the DOS at $E_F$ obtained from photoemission spectra at low temperature depends critically on the accuracy of the correction of the surface photovoltage effect. The first indication for a gap in Si(111)-(7x7) came from the high-resolution electron-energy-loss-spectroscopy data taken at 20 K by Persson and Demuth [8] which was interpreted with a model that assumes a gap of ~40 meV and a very narrow DOS peak at $E_F$. A second indication came from the β-NMR data between 50 K and 300 K of Ref. [9] which was interpreted with a model that assumes a surface gap between 100 meV and 500 meV and a narrow band at $E_F$ with a width of ~10 meV. This narrow band is also present in the results of theoretical calculations of the Si(111)-(7x7) that include electron correlation effects [15]. The deep minimum at $E_F$ in our spectra above 20 K and the gap below 15-20 K agrees with the gaps resulting from the above models, but we do not observe the narrow DOS peak at $E_F$ even though our energy resolution is better than 10 meV.

The electric transport properties of the Si(111)-(7x7) surface have been measured by the method of micro-four-point probes at 100 K [11]. The measured value of the surface conductivity was found to be "below the lowest reasonable value for metallic conductivity", and therefore "consistent with the hypothesis that Si(111)-(7x7) could be on the verge of a metal to nonmetal transition" [11]. Our STS measurements indicate that most of the drop of the differential conductance at $E_F$ occurs below 60 K, but also above this temperature the electrical conductivity could be low because of electron-phonon and/or electron-electron interactions that cause the low-temperature insulating state.

Finally the question is, whether the observed gap is primarily due to electron-electron interactions and thus resulting from a Mott-Hubbard mechanism, or whether it is dominated by electron-phonon interaction and thus connected with lattice distortions. According to the local density approximation calculations of Ref. [15], there are several adatom-derived bands near $E_f$ accommodating an odd number of electrons. If only one of these bands crosses $E_f$, and therefore is half-filled, either a structural distortion mechanism or a Mott-Hubbard mechanism could cause the opening of the gap. If two bands cross $E_f$, there is no half-filling and the Mott-Hubbard mechanism alone cannot work. In this case the lattice distortions should open two gaps at the two Fermi surfaces or split the two bands far apart and open a gap in the now half-filled upper band. A possible scenario including a combination of both effects is the following: a distortion splits the unfilled bands and the electron-electron interaction opens a gap in the half-filled upper band. Because the electron-phonon coupling constant is large according to Ref. [7], and because Fig. 2 of Ref.

[15] indicates that two bands cross $E_f$, it is likely that the ground state of the Si(111)-(7x7) has a small lattice distortion not yet reported.

In conclusion we have tested a method for the correction of systematic errors in the STS spectra caused by the transport of the injected carriers. The method works when the pinning of the Fermi level at the surface is efficient and should be applicable to many wide gap semiconductors and to samples that do not form ohmic contacts with the sample holder. We have verified the presence of an energy gap in the ground state of the Si(111)-(7x7) surface below ~20 K and have measured it. The mechanism that opens the gap should involve a small distortion of the (7x7) unit cell with or without a Mott-Hubbard transition. STM below 20 K should be able to identify such a distortion if it is static.

We acknowledge financial support from the Sonderforschungsbereich 508 "Quantenmaterialien" and from the Graduiertenkolleg 1286 "Functional Metal-Semiconductor Hybrid Systems" of the Deutsche Forschungsgemeinschaft. This work was also supported by the MIUR-PRIN (20087NX9Y7). We thank F. Meier and L. Zhou for assistance during measurements, and E. Tosatti and G. Profeta for many discussions.

**Figure Captions:**

Fig. 1.
Sketch of the profile of the potential V on the path from the sample holder to the STM tip at large total resistance $R_{total}$ – i.e. large tip-surface distance- (a) and at small $R_{total}$ –i.e. small tip-surface distance- (b) for an n-type semiconductor sample, negative bias applied to the sample holder, and the same current flowing through the system in both cases. An efficient pinning of the Fermi level at the surface is assumed. The potential drop between the semiconductor surface below the tip and the sample holder $V_{surf-holder}$ is caused by the metal-semiconductor junction between the sample and the holder, the pinning of the Fermi level at the surface is caused by the surface states, and by the flow of the current. It depends on the current and does not depend on the tip-sample distance. The bias voltage at small $R_{total}$ -$V^s_{bias}$ - is approximately $V_{surf-holder}$. Panel (c) is a sketch of the energy levels as a function of the position from the holder to the tip for large $R_{total}$ and small $R_{total}$, $E_C$ and $E_V$ are the bottom of the conduction band and the top of the valence band respectively, SS are surface states. An approximate equivalent circuit is given by a diode at the holder-sample junction, a resistor in the semiconductor bulk, a second diode in the surface region, and a second resistor in the vacuum barrier.

Fig. 2
Relation between the voltage $V_{bias}$ applied between the STM tip and the sample holder and the tunnelling current I for different stabilization resistances $R_{total}$ on Si(111)-(7x7) at 0.3 K. The stabilization voltage $V_{stab}$ was +1.6 V. At a given value of I the potential drop $V_{tip-surf}$ between the tip and the surface at high $R_{total}$ can be approximated by the difference between $V_{bias}$ and the $V_{bias}$ measured with much lower $R_{total}$, i.e. much higher stabilization currents. At low $R_{total}$ $V_{bias}$ approximates the potential drop $V_{surf-holder}$ between the surface and the sample holder.

Fig. 3
Differential conductance of the Si(111)-(7x7) surface at 0.3 K for different $R_{total}$ at the stabilization voltage of +1.6 V. The spectra are the average of 200 spectra measured in the region shown in the inset constant current image (V=1.6 V, I=200 pA). The non-rigid shifts of the features of the spectra for decreasing $R_{total}$ point to effects caused by the charge transport between the surface and the sample holder.

Fig. 4
(a) Corrected normalized conductance $(dI/dV_{tip-surf})(I/V_{tip-surf})$ of the Si surface at 0.3 K for two different $R_{total}$. The arrows point to the DOS features of the rest atoms (R) and of the adatoms (A). (b) Corrected differential conductance near the gap region at 0.3 K for $R_{total}$=130 GΩ. The straight lines are used to estimate the gap by a linear extrapolation of the near gap spectrum. (c) Normalized conductance $dI/dV_{bias}/(I/V_{bias})$ of Si samples with improved electric contacts to the sample holder measured at 295 K and 5K with a stabilization resistance of $R_{total}$=80 GΩ at a stabilization voltage of $V_{stab}$=-1.6 V. The 295 K spectrum is vertically displaced by +1.5. (d) Differential conductance $dI/dV_{bias}$ at 5 K and 295 K and the linear fit used to estimate the gap at 5K.

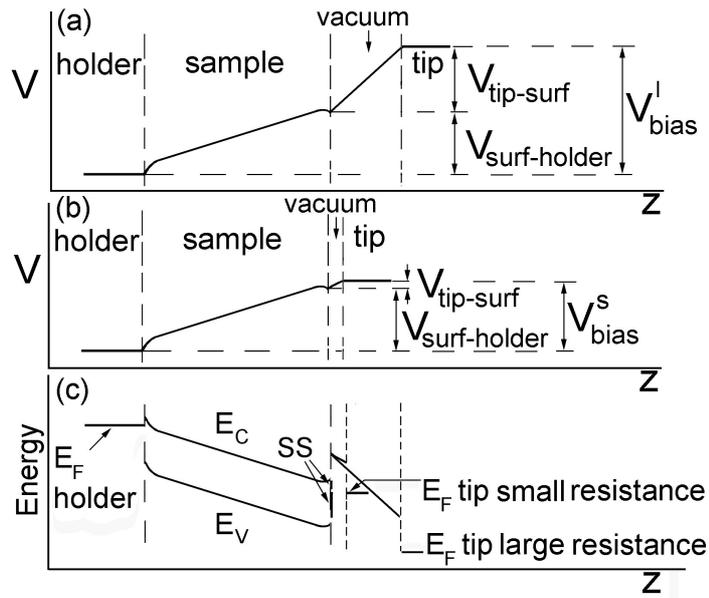

Figure 1

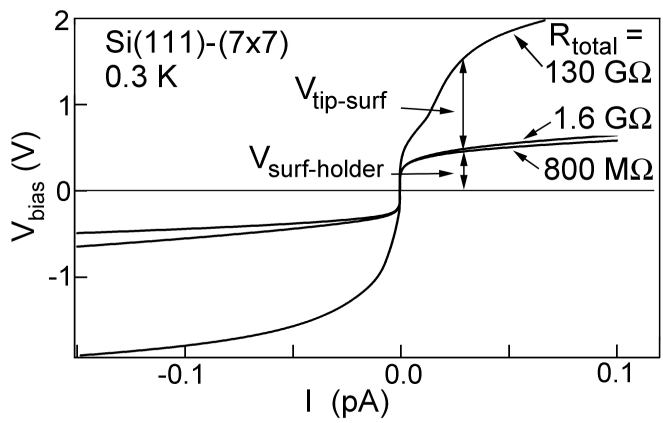

Fig. 2

Figure 3

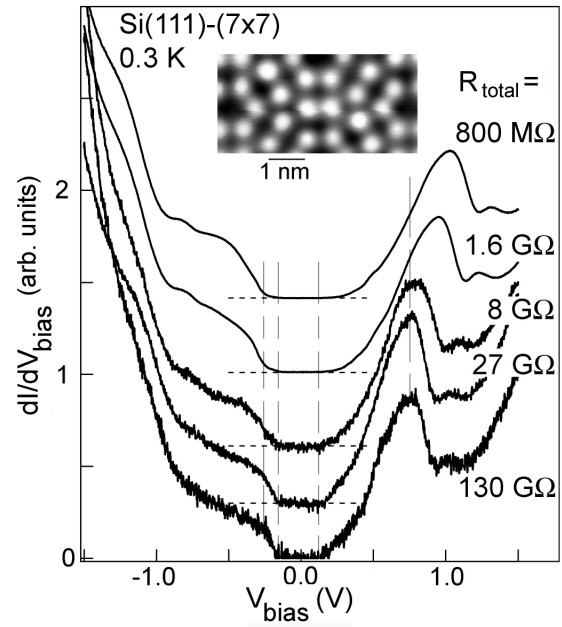

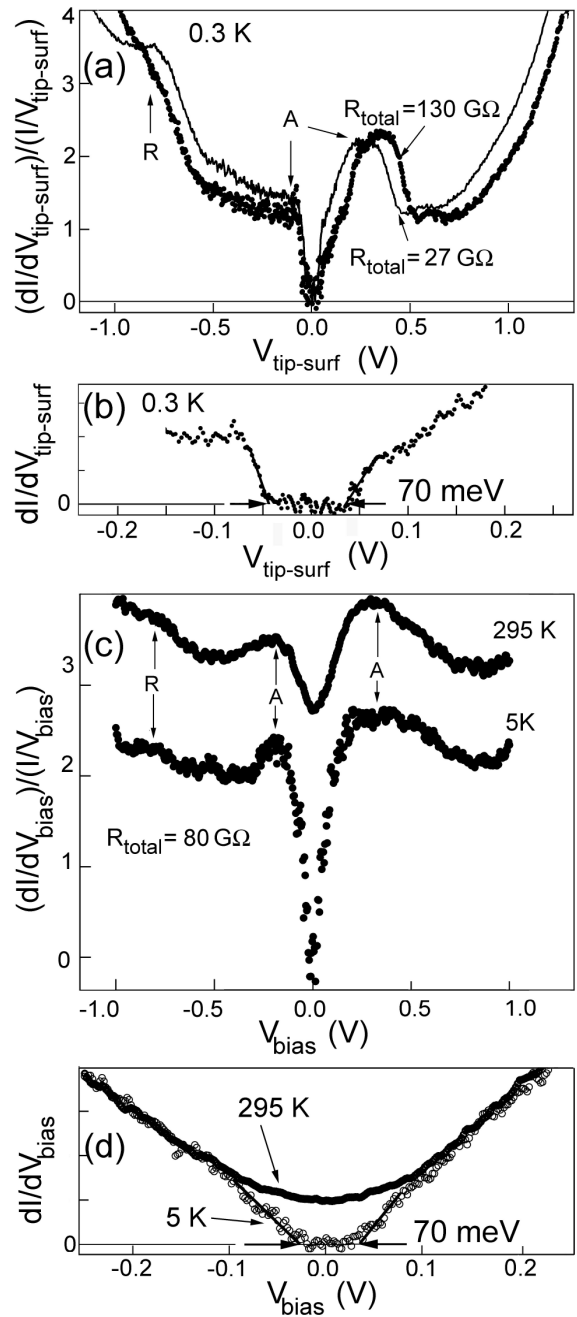

Fig. 4